\documentclass[prl,aps,superscriptaddress,notitlepage]{revtex4}
\usepackage{bm}
\usepackage[usenames,dvipsnames]{xcolor}
\usepackage[breaklinks,bookmarks = false,pdfpagemode = UseNone, colorlinks= true,]{hyperref}
\hypersetup{linkcolor=RubineRed,citecolor=RoyalBlue,filecolor=Mulberry,urlcolor=RoyalBlue}
\usepackage{times}
\usepackage{latexsym}
\usepackage{graphicx}
\usepackage{microtype}

\newcommand{\NNO}{NdNiO$_3$}
\newcommand{\jwf}[1]{\textcolor{Black}{#1}}

\begin{document}

\title{Magnetic Order Driven Ultrafast Phase Transition in NdNiO$_3$}

\author{V.A. Stoica}
\affiliation{%
Department of Materials Science and Engineering, Pennsylvania State University, University Park, Pennsylvania 16802, USA}
\author{D.\ Puggioni}
\affiliation{%
Department of Materials Science and Engineering, Northwestern University, Evanston, Illinois 60208, USA}
\author{J.\ Zhang}
\affiliation{%
Department of Physics, The University of California at San Diego, La Jolla, California 92093, USA}
\author{R.\ Singla}
\affiliation{%
Department of Physics, The University of California at San Diego, La Jolla, California 92093, USA}
\author{G.L. Dakovski}
\affiliation{%
Linac Coherent Light Source, SLAC National Accelerator Laboratory, 2575 Sand Hill Road, Menlo Park, CA 94025, USA}
\author{G. Coslovich}
\affiliation{%
Linac Coherent Light Source, SLAC National Accelerator Laboratory, 2575 Sand Hill Road, Menlo Park, CA 94025, USA}
\author{M.H. Seaberg}
\affiliation{%
Linac Coherent Light Source, SLAC National Accelerator Laboratory, 2575 Sand Hill Road, Menlo Park, CA 94025, USA}
\author{M.\ Kareev}
\affiliation{%
Department of Physics and Astronomy, Rutgers University, Piscataway, New Jersey 08854, USA}
\author{S.\ Middey}
\affiliation{%
Department of Physics and Astronomy, Rutgers University, Piscataway, New Jersey 08854, USA}
\author{P.\ Kissin}
\affiliation{%
Department of Physics, The University of California at San Diego, La Jolla, California 92093, USA}
\author{R.D.\ Averitt}
\affiliation{%
Department of Physics, The University of California at San Diego, La Jolla, California 92093, USA}
\author{J.\ Chakhalian}
\affiliation{%
Department of Physics and Astronomy, Rutgers University, Piscataway, New Jersey 08854, USA}
\author{H.\ Wen}
\affiliation{%
Advanced Photon Source, Argonne National Laboratory, Argonne, Illinois 60439, USA}
\affiliation{%
Materials Science Division, Argonne National Laboratory, Argonne, Illinois 60439, USA}
\author{J.M.\ Rondinelli}
\affiliation{%
Department of Materials Science and Engineering, Northwestern University, Evanston, Illinois 60208, USA}
\author{J. W.\ Freeland}
\email{freeland@anl.gov}
\affiliation{%
Advanced Photon Source, Argonne National Laboratory, Argonne, Illinois 60439, USA}

\begin{abstract}
Ultrashort x-ray pulses can be used to disentangle magnetic and structural dynamics and are accordingly utilized here to study the photoexcitation of \NNO (NNO), a model nickelate exhibiting structural and magnetic dynamics that conspire to induce an IMT. During the course of the photoinduced insulator to metal transition (IMT) with above gap excitation, we observe an ultrafast ($<$ 180 fs) quenching of magnetic order followed by a time delayed collapse of the insulating phase probed by X-ray absorption and THz transmission (~ 450 fs) that correlates with the slowest optical phonon mode involved in the structural transition. A simultaneous order-disorder response at the Ni site and displacive response at the Nd site coexist in the ultrafast magnetic response. Crucially, we observe the optical phonon through its coherent coupling with Nd magnetic order, demonstrating that the magnetic and structural degrees of freedom both contribute in driving the IMT. Density functional theory (DFT) calculations reveal a consistent scenario where optically driven inter-site charge transfer (ICT) drives a collapse of antiferromagnetic order that in turn destabilizes the charge-ordered phase resulting in an IMT. These experiments provide new modalities for control of electronic phase transitions in quantum materials based on ultrafast interplay between structural and magnetic orders created by femtosecond photoexcitation.
\end{abstract}

\date{\today}
\maketitle

\section{Introduction}

Non-conventional superconductivity (SC) and  insulator-metal transitions (IMT) in transition metal oxides (TMO) are topics at the forefront of research in condensed matter physics because their mechanisms are not fully understood. Therein, microscopic interactions between structural, electronic, orbital and magnetic degrees of freedom (DOF) conspire to generate macroscopic quantum phases and large changes in properties associated with transitions between different states. Quite often, however, these DOF are deeply intertwined, leading to competing, coexisting, or even cooperative orders \cite{Tokura:2000ck,Dagotto:2005ip,Keimer:2017gq}. Such cooperative coupling between the lattice structure, electronic order, and the magnetic configuration occurs in a wide variety of transition metal oxides such as manganites\cite{Tokura:2006ff}, cobaltites\cite{Sundaram:2009bx}, ferrites\cite{Blasco:2018ep}, vanadates\cite{Yan:2019ft}, nickelates\cite{Catalano:2018kc,Li:2019jy}, and cuprates\cite{Fradkin:2015co}. This arises due to the highly connected nature of the lattice\cite{Woodward:1997th,Howard:1998uf,Howard:2004co,Carpenter:2009cf} that results in an interplay between changes in electronic and magnetic order with many structural degrees of freedom such as distortions, octahedral rotations and cation displacements. Recently, some data has been assembled and analyzed to explore correlations between structural, electronic, and magnetic order\cite{Balachandran:2013cg,Wagner:2016dx,Wagner:2018kz}, which is insightful, but not always conclusive. It is thus challenging to disentangle the role of different DOF. especially in the case where multiple order parameters change under the same physical conditions. Here, we focus on magneto-structural interactions in TMOs that can dictate the realization and control of macroscopic quantum phases\cite{Imada:1998er}.
\begin{figure}[h]
\centering
\includegraphics[width=.35\textwidth]{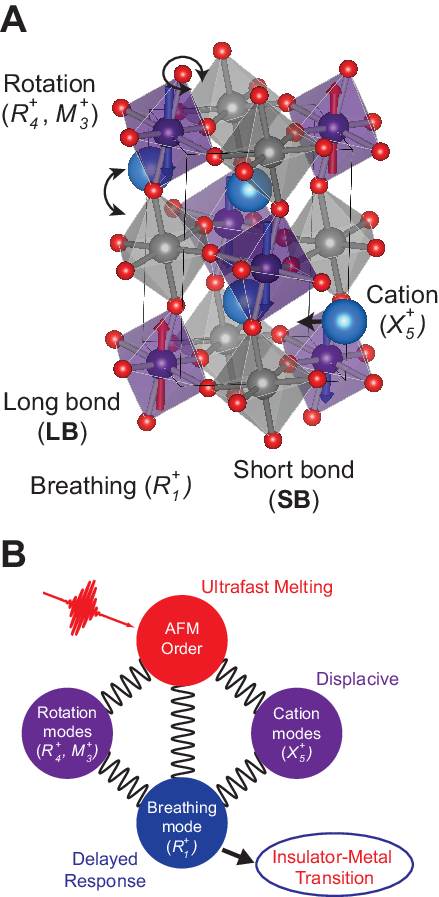} 
\caption{\label{Fig1} {\bf Coupled order parameters} (A) NdNiO$_3$ monoclinic structure highlighting different structure, electronic and magnetic degrees of freedom. Octahedral sites connected with the (purple) long bond (LB) and (grey) short bond (SB) charge order associated with the Ni-O breathing mode distortion. Part of the E$^\prime$-type antiferromagnetic (AFM) unit cell  is shown with alternating planes of ferromagnetically aligned $S=1$ LB site (purple) and $S=0$ SB sites (grey) that form the $\cdots\uparrow\cdot\ 0\ \cdot\downarrow\cdot\ 0\ \cdots$ pattern (B) An overview of how these intertwined orders are connected in \NNO.
}
\end{figure}

To try to disentangle these subtle interactions, static-tuning knobs such as strain, composition, electrostatic gating, or magnetic fields are most often employed to explore intricate energy landscapes with the possibility of selecting a particular ordered quantum phase\cite{Tokura:2017bh}. Nevertheless, the interaction between DOF is dynamic in nature and occurs at fundamental ultrashort timescales that need to be accessed directly in order to sort out the basic physics. In this way, the dynamic exploration of the energy landscape can follow how phases evolve in real time during the conversion between quantum states. Furthermore, this route can make it possible to disentangle how microscopic competing degrees of freedom lead to the emergence of long range order \cite{Averitt:2002dq,Basov:2011ht,Zhang:2014dq,Giannetti:2016hp}, with the ultimate goal of light directed property control \cite{Basov:2017ix}. Connected with this goal, ultrafast techniques now span the electromagnetic spectrum enabling multi-modal studies of complexity in solids with X-ray techniques having risen to prominence for quantitative probing of these DOF \cite{Lindenberg:2000ib,Cavalleri:2005fe,Ichikawa:2011js,Lee:2012ga,Caviglia:2013to,deJong:2013he,Park:2013jt,Beaud:2014bp,Forst:2015fv,Langner:2015ef,Lourembam:2015kb,Zhu:2016cp,ThielemannKuhn:2017cj}. Indeed, such ultrafast structural techniques have been combined with THz probes to reveal the evolution of structural and electronic dynamics during conductive phase transitions that are not accessible in steady state\cite{Morrison:2014cs,Forst:2015fv,Wall:2018eg,Otto:2019ki,Hu:2014cg,Forst:2017ip}. For example, during the photoinduced IMT in VO$_2$ an unexpected order-disorder character of the transition was discovered\cite{Morrison:2014cs,Wall:2018eg}, where a sub-ps disordering of monoclinic ground state is followed by a slower growth of rutile metallic phase in a few ps, both contributing to the electronic response\cite{Otto:2019ki}. Resonant phonon excitation has also revealed the coupling of structure and electronic DOF in a cuprate SC\cite{Hu:2014cg}, where resonant excitation of c-axis motion of apical oxygen atoms triggered an enhanced SC tunneling, suggesting possible avenues for targeted structural tuning of phase transitions. 

In this article, we focus on the perovskite nickelate NdNiO$_3$ as a prototypical system with coupled order parameters, exhibiting concomitant charge and magnetic order associated with an IMT (Refs.\ \onlinecite{Freeland:2015iw,Middey:2016jc,Catalano:2018kc} and references therein). Charge order is associated with an orthorhombic to monoclinic structural transition involving two NiO$_6$ sites in the monoclinic phase, which are referred to as short-bond (SB) and long-bond (LB) (see Fig.\ \ref{Fig1}A ).  The magnetic order is E$^\prime$-type antiferromagnetic (AFM) with a 4$\times$4$\times$4 pseudocubic unit cell (2$\times$1$\times$2 monoclinic unit cell) with large planes of ferromagnetically aligned LB ($S=1$) and small SB ($S=0$) Ni sites arranged in an $\cdots\uparrow\cdot\ 0\ \cdot\downarrow\cdot\ 0\ \cdots$ 
pattern, part of which is shown in CO unit cell in Fig.\ \ref{Fig1}A \cite{GarciaMunoz:1992dj,RodriguezCarvajal:1998dy, Scagnoli:2006ja}.  Although the nominal ionic ground-state is Ni$^{3+}$ in a low spin 3d$^7$ configuration ($t_{2g}^6e_g^1$), theory strongly supports a state that is $3d^8\underline{L}$ where $\underline{L}$ denotes a ligand hole on the oxygen site. In this scenario, the charge ordered phase corresponds to alternating $3d^8$\,LB and $3d^8\underline{L}^2$\,SB sites \cite{Mizokawa:2000wq,Mazin:2007jx,Lee:2011eg,Park:2012hg,Johnston:2014ca,Subedi:2015en,Varignon:2017is,Haule:2017ft}. For \NNO, theory suggests that the magnetic order could contribute to the stability of the CO phase \cite{Park:2012hg,Varignon:2017is,Mercy:2017il}, but this has been difficult to verify experimentally and motivates our time domain studies.

Upon resonant phonon (mid-infrared) pumping  of charge and magnetic ordered NdNiO$_3$\cite{Forst:2015fv,Forst:2017ip}, the dynamic evolution of magnetism was observed and compared with THz conductivity changes, indicating supersonic magnetic front propagation during IMT. However, similar ultrafast conductivity dynamics are also observed in SmNiO$_3$ above and below the magnetic transition temperature\cite{Hu:2016hha}, indicating that magnetism does not play a decisive role in driving the ultrafast IMT explored with mid-infrared pumping. In contrast, static strain and dimensionality tuning of NdNiO$_3$ heterostructures\cite{Meyers:2016fc} implicate magnetic ordering as the origin for the IMT since charge ordering and structural symmetry were suppressed, providing support for Mott physics in this system with potential for ultrafast magnetic control of the conductivity. Additionally, the conductivity changes are usually observed to be much larger in PrNiO$_3$\cite{Hepting:2014ip} and NdNiO$_3$\cite{Catalano:2015ff}, exhibiting simultaneous magnetic and charge order at IMT, compared to other RNiO$_3$ (R - rare earth element) compounds, where magnetism orders at lower temperatures\cite{Catalano:2018kc}, raising the question on how magnetism is enhancing the conductivity at the transition. This calls for selective control of degrees of freedom using ultrafast excitation to establish their role in the IMT. Thus, RNiO$_3$ represent an insightful platform for ultrafast studies, where the dynamic interplay between electronic, magnetic and structural DOF can inform the pathway toward the IMT.

\jwf{To address these questions, we harness ultrafast soft X-ray scattering and absorption capabilities at the Linac Coherent Light Source (LCLS) to probe the electronic and magnetic degrees of freedom directly in order  to disentangle multiple interactions in a correlated oxide. In the equilibrium phase diagram, NdNiO$_3$ has combined antiferromagnetic (AFM) and charge order (CO) that collapse simultaneously at the IMT.  Using time-resolved magnetic scattering  sensitive to the long-range spin order, XAS sensitive to the local Ni coordination connected to the breathing mode distortion of the charge ordered (CO) state, and THz transmission as a probe of carriers, we find that for above gap excitation, there is a rapid collapse of the magnetic state ($\leq$ 175 fs) followed by a slower relaxation of the breathing mode CO and IMT response ($\sim$ 450 fs). From comparison between experiment and theory, we  develop a clear picture of the pathway where an inter-site charge transfer (ICT)  triggers an ultrafast collapse of the magnetic state followed by a slower IMT. The IMT is triggered by simultaneous collapse of magnetic order and the displacive excitation of a coherent Nd-O soft phonon that is seen to coherently perturb the Nd magnetic order dynamics, in direct support of spin-phonon coupling\cite{Hong:2012dc}. We highlight the pathway for the optically driven transition in Fig.\ \ref{Fig1}B showing the interconnection between the different degrees of freedom. We will refer here to the leading distortion modes\cite{Balachandran:2013cg,Wagner:2018kz} that are relevant for the monoclinic to orthorhombic (M-O) symmetry change that is always observed at the IMT in the structural refinements at equilibrium in RNiO$_3$ compounds to explore their role in our experiments. A depiction of the IMT transition pathway starting from ultrafast melting of AFM order that ultimately triggers the collapse of charge-order is shown in Fig.\ \ref{Fig1}B. This process is slowed by the coupled lattice dynamics involving octahedral rotation changes ($R^{+}_4$ and $M^{+}_3$ distortions), and anti-polar A-site cations displacements ($X^{+}_5$) that are coupled to the breathing mode ($R^{+}_1$) and ultimately dictate the response time of the IMT. Our results provide a direct visualization of the distinct electronic and magnetic dynamics of the IMT, which was not accessible under steady state equilibrium interrogation.}

\begin{figure}
\centering 
\includegraphics[width=.45\textwidth]{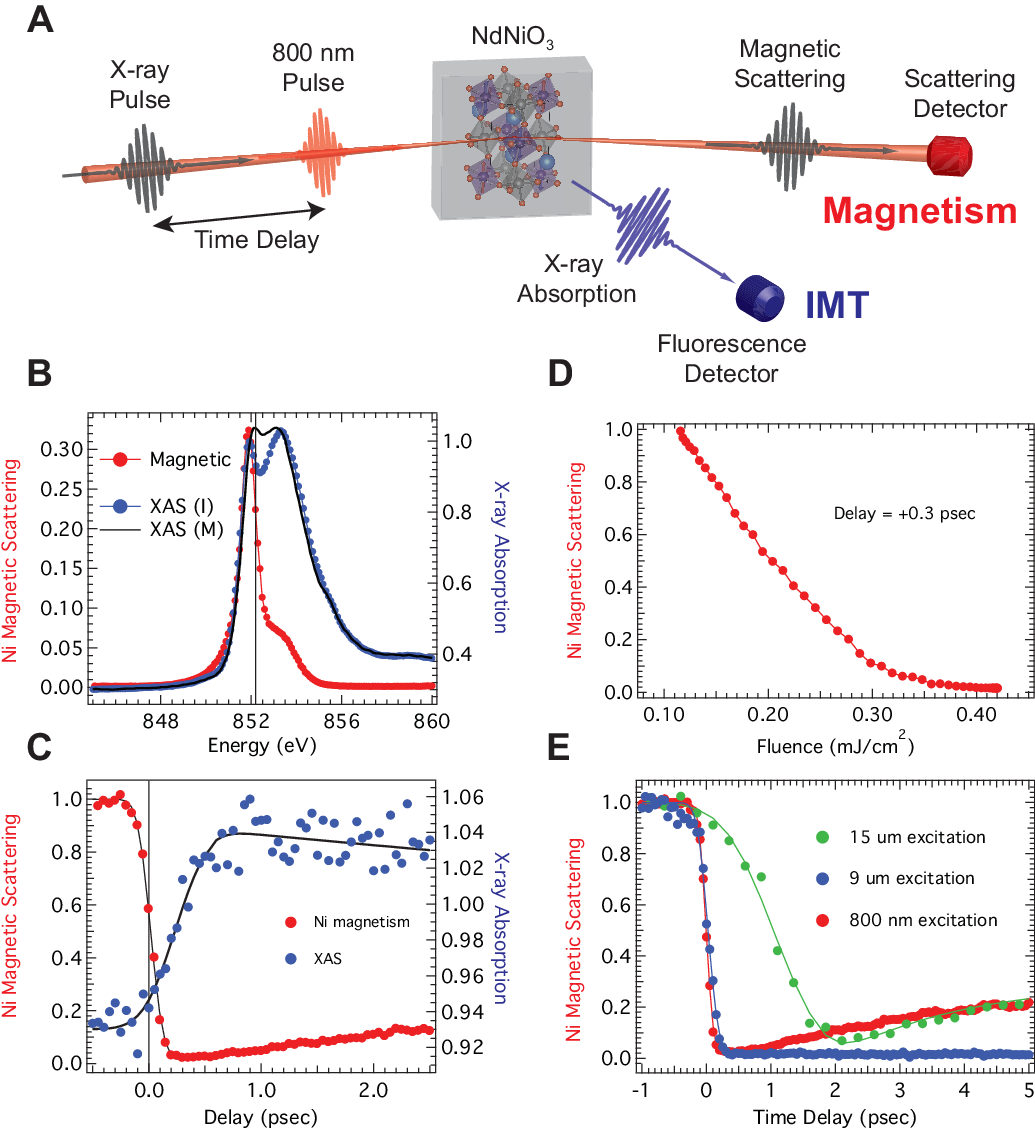} 
\caption{\label{Fig2} {\bf Experiment geometry and }  (A) Illustration of sample geometry and experiment. (B) Resonant X-ray  magnetic scattering spectra at the Ni L$_3$ edge and absorption above and below the IMT. The vertical line marks the energy for the measurement of the time delay scans. (C) Delay scans for the magnetic scattering and XAS data following 800 nm excitation. The scale for the magnetic scattering is 1 for static AFM order and zero for complete loss of long-range order. XAS is a relative change consistent with the magnitude of XAS change seen with crossing IMT with temperature. (D) Fluence dependence of the Ni magnetic scattering. (E) }
\end{figure}

\section{Experimental Approach}

The (001) oriented films were grown using pulsed laser deposition on NdGaO$_3$ substrates (tensile strain of 1.4 $\%$)  with a thickness of 50 nm \cite{Liu:2013ep,Middey:2016jc}. The thickness was chosen to show bulk-like properties and be matched with the pump/probe absorption length while avoiding complications from dimensionality and orientation that can lead to an altered ground-state\cite{Middey:2016jc,Catalano:2018kc}. Experiments were performed at the SXR end-station where the X-rays and laser pulses arrive colinearly with the polarization in the scattering plane (see Fig.\ \ref{Fig2}A)\cite{Schlotter:2012ey}. The sample was cooled below the IMT temperature (T$_{IMT} \approx$ 150\,K) to 70\,K and aligned to measure the off-specular $(1/4,1/4,1/4)$ pseudocubic Bragg peak corresponding to the AFM-E$^\prime$ order\cite{Scagnoli:2006ja,Scagnoli:2008iu} as well as the bulk-sensitive X-ray absorption spectroscopy (XAS) in fluorescence yield (see Fig.\ \ref{Fig2}B), where the vertical line shows the energy chosen for the time-resolved  measurements. Since the main change in the XAS across the IMT manifests at the dip near 852.5 eV, as shown by comparing the XAS in the insulating (I) vs metallic phase (M) in Fig.\ \ref{Fig2}B, this energy was chosen to measure both channels without changing the photon energy. The pulse duration for this experiment was $\sim$100\,fs for the X-ray pulses and $\sim$150\,fs for the laser, which gives a time resolution of $\sim$175\,fs.  Excitation pulses at 1.55\,eV were utilized  in addition to mid-IR pulses generated using optical parametric generation and difference frequency generation ($\approx$ 83 - 135 meV) to enable above- and below-gap pumping (optical gap E$_{g} \approx $ 100meV) \cite{Dhaka:2015id}. Additionally, the insulator-to-metal transition dynamics of the \NNO thin film was measured with optical pump-THz spectroscopy probe to track the formation of the metallic phase. To achieve optimal temporal resolution, the experiment was implemented in transmission with both excitation
(optical) and probe (THz) beams being collinear and perpendicular the sample to avoid temporal broadening at oblique incidence for the THz probe beam. With normal incidence geometry, we were able to achieve a temporal resolution of 150 fs as determined by a reference transient THz transmission measurement on a GaAs crystal.

\section{Ultrafast phase transitions}
\jwf{To set the stage for understanding ultrafast magnetic and electronic data below, we first summarize recent work looking at both the excitation pathway and what controls the fundamental timescales for collapse of these different degrees of freedom. As noted in the reviews\cite{Averitt:2002dq,Basov:2011ht,Zhang:2014dq,Giannetti:2016hp}, there has been an extensive amount of work on dynamics in quantum materials. Here we want to focus on two aspects. First is understanding the timescale of the transition. For example, a recent study of orbital order in manganites connects the transition time to specific Jahn-Teller lattice modes related to the quenching of long-range orbital order\cite{Singla:2013kk,Beaud:2014bp}. In the case of ultrafast magnetism\cite{Kirilyuk:2010ha}, we need to clarify some distinctions for magnetic materials studied in magnetic experiments. The first is optical control pathways in metallic vs.\ insulating magnets can be quite different. In metallic systems, the thermalization of hot carriers plays a very important role whereas in insulators the process is connected to how photoexcitations alter the magnetic state. For the insulating case of \NNO we consider the main pathways to be related to direct coupling to magnon excitation via optical charge-transfer that allows the light to couple to the magnetic degree of freedom\cite{Hellsvik:2016cs}.The second aspect is that unlike ferromagnetic/ferrimagnetic systems, the change in net angular momentum is zero since both the ordered and disordered state have zero angular momentum. Another aspect of the transition is the thermal energy generated by laser excitation. A simplistic  estimation for the photoinduced transient heating can be made based on the known specific heat\cite{Hooda:2016bj}, our calculated optical absorption in the experiment (0.4) and optical penetration depth of 50 nm\cite{Ruello:2009ew}, giving a transient temperature of 110 K at the peak fluence at saturation, which falls well below T$_{IMT}$ and indicates its non-thermal character. }

\section{Tracking Ni Magnetism}
\jwf{By tuning to the AFM order peak at the Ni resonance, the ultrafast magnetic dynamics measured by pump-probe delay scans following 1.55 eV excitation are presented in Fig.\ \ref{Fig2}C.  At an excitation level of $\sim$0.01 electrons per nickel site (0.5 mJ/cm$^2$), the magnetic scattering is completely quenched within 175 fs.  The transition times for all delay scans were determined by fitting to an error function together with a slow exponential recovery. Note that the quoted transition times are defined by the twice the Gaussian width (width defined by difference at 0.85 and 0.15 of maximum value) and fitting errors were all $<$ 10 fs. As noted above, given the experimental resolution, the magnetic order collapse is at the limit of our temporal resolution. The fluence dependence at early delay times is shown in Fig.\ \ref{Fig2}D, which displays a clear curve saturating   close to 0.4 mJ/cm$^2$. We note here that compared to other optically driven phase transitions in vanadates\cite{Cavalleri:2004eh,Wall:2013dv}, manganites\cite{Beaud:2014bp},cuprates\cite{Johnson:2012di}, nickelates\cite{Lee:2012ga,Beyerlein:2020be}, \NNO\ has a fluence threshold that is nearly an order of magnitude smaller. This indicates a more efficient coupling of the optical pump to control magnetic order. }

\jwf{Another important consideration is the role of the pump photon energy in terms of controlling the dynamics. Figure\ \ref{Fig2}E highlights that the timescale of the magnetic collapse remains sub-200 fs for above bandgap excitations, where the bandgap of NdNiO$_3$ has been shown to be $\sim$100 meV\cite{Katsufuji:1995dy,Medarde:1997bu,Okazaki:2003to}. This is evident for 800 nm (1.55 eV, red curve), and 9 $\mu$m (138 meV, blue curve). Importantly, this data also reveals that the excess kinetic energy imparted to the carriers with 1.55\,eV photons does not play a crucial role in the dynamics given the similar timescale with 138 meV excitation.  In contrast, for sub-gap excitation with 15 $\mu$m (83 meV, green curve), the magnetic transition is considerably slower  (1.2 ps). Note, however, that the incident fluence level for the 15 $\mu$m pumping required to quench magnetic order is 4 mJ/cm$^2$ , while the 9 $\mu$m data were  acquired at 10 mJ/cm$^2$. The increase in fluence is likely related to the different reflectivity and longer penetration depth for longer wavelength photons. For the 15 $\mu$m case below the gap,  charge carrier excitation pathway is absent and the light couples to the lattice (phonons) to drive the collapse of the magnetism  \cite{Caviglia:2012ea,Forst:2015fv,Forst:2017ip}.  This data clearly reveals two distinct pathways leading to a collapse of the magnetic order. Namely, with below-gap phonon excitation there is a slower pathway, but above-gap pumping of charge carriers leads to a far faster non-thermal route where by the electronic excitation triggers the subsequent IMT dynamics. In the following, we will focus on the faster above gap route since the lattice excitation has been previously discussed in detail\cite{Caviglia:2012ea,Forst:2015fv,Forst:2017ip}. }

\jwf{With an understanding of the timescale- and pump-dependence, we can construct a more complete picture of how light interacts with these degrees of freedom.  First, we discuss the nature of the optical excitation process. In the case of highly-covalent nickelates, our calculations of the optical spectra (see supplement and those of Refs. \cite{Ruppen:2017du,Bieder:2020bz} ) are consistent with an inter-site charge transfer (ICT) involving excitations between the Ni $3d$ and O $2p$ states which changes the charge distribution around the  $3d^8$ LB sites to  $3d^8\underline{L}^2$ SB sites. As we will show in the theory section below, the change in the charge density, primarily on the Ligand hole states, changes both the nature of the magnetic exchange and the magnetic moment on the SB site. In the AFM phase, the SB site is in a S=0 state, which can be stabilized under the crystal field from to the high Ligand hole density of the $3d^8\underline{L}^2$ SB site\cite{VanDerLaan:1988jp}.  With the change in Ligand hole density driven by optical d-p excitations, this changes the LB-SB exchange and results in a non-zero moment on SB Ni site due to an orbital rearrangement. As such, we consider ICT an operative pathway by which changes in the local electronic and magnetic configuration can modify the magnetic order via optical modification of the exchange interactions. Consequently, the timescale for the collapse of magnetism can then be tied to the details of the spin-wave spectrum as well as the time-scale of the ICT.  Furthermore, the lattice excitation at the Nd site proceeds coherently and in phase (displacive) with this event, as we will show later. For other complex oxide systems, the timescales for the charge transfer excitation and rearrangement of orbital occupancies has been shown to occur on a $<$100 fs timescales\cite{Okamoto:2011da,Singla:2013kk,Beaud:2014bp}. The spin-wave generation in other insulating AFM (KNiF$_3$) showed that the transition dynamics involves zone-boundary spinwaves, due to the momentum conserving optical excitation of bimagnons\cite{Hellsvik:2016cs,Bossini:2016iv}. In the case of \NNO, recent resonant inelastic X-ray scattering measurements show that the zone-boundary magnons in NdNiO$_3$ have $\sim$50 meV energy($\sim$85 fs period)\cite{Lu:2018cf}. Given the sub-100 fs timescales for both parts of the process, we conclude that the true Ni AFM order collapse likely occurs at sub-100 fs timescales.}

\begin{figure}[t]
\centering
\includegraphics[width=.45\textwidth]{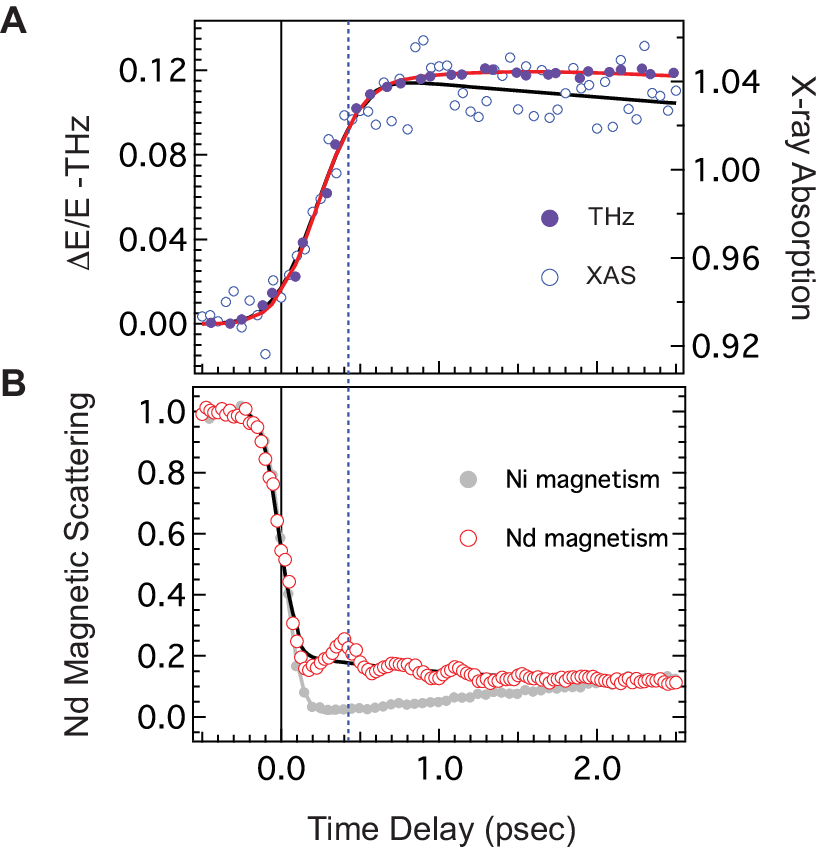} 
\caption{\label{Fig3} {\bf Magnetic and electronic dynamics}  (A) THz transmission overlayed with XAS to confirm the timescale of the IMT. (B) Nd magnetic scattering intensity at the same wave-vector as Ni. The thin vertical line is at time-zero while the second dashed line is after 1 period of the coherent phonon.}
\end{figure}

\section{Insulator-Metal Transition Dynamics}
\jwf{To understand the implications of magnetism collapse, we now turn to the changes in electronic properties and how they are connected to the magnetism. First we consider the changes in X-ray absorption spectra (XAS) in the insulator vs.\ metallic phase as shown in Fig.\ \ref{Fig2}B. Previous work and calculations have shown how the formation of the insulating charge ordered (CO) state is correlated to changes in the absorption, which is not influenced by the magnetic order \cite{Freeland:2015iw,Green:2016dk}. For the delay scan shown in Fig.\ \ref{Fig3}A, we have set the photon energy to the vertical line in Fig.\ \ref{Fig2}B, where there is $\sim$10\% change in the XAS across the IMT. In dramatic contrast to the magnetic dynamics, the time-resolved XAS scan shows a longer transition time of 446 fs, which demonstrates that the magnetic and electronic contributions to the IMT have different timescales. The rapid  and total collapse of the magnetic order occurs prior to changes in the CO (breathing distortion) and IMT embodied in the XAS response. To confirm that the changes in XAS are tied to the IMT, we utilized optical-pump THz-probe measurements to directly follow the formation of the metallic state. As shown in Fig.\ \ref{Fig2}B, the THz response shows a transition to the metallic phase that correlates directly with the XAS signal, where a transition time to the metallic phase from the THz transmission is 425 
fs. The IMT transition time is also consistent with recent time-resolved measurements that measured the collapse of the CO state\cite{Esposito:2018kb}, indicating that the IMT is tracking the loss of CO and not the magnetism. }
 
\jwf{For bulk RNiO$_3$ compounds, it is known that the paramagnetic phase can exist in the presence of CO in the cases of small R cation size, connected to the fact that a change in lattice symmetry is not required for the collapse of magnetic order. However, the metallic state is only present in orthorhombic symmetry. On the structural side, the collapse of the CO is key to the transition to the higher symmetry metallic phase, while the breathing mode is also coupled to a low frequency A-site cation phonon mode that is directly associated with the $P2_1/n \rightarrow Pnma$ structural phase transition\cite{Balachandran:2013cg}. From analyses of existing structural refinement data at steady state, we find a linear relationship between the $X^{+}_5$ distortion that is characteristic for A-site cation distortion and the breathing mode  ($R^{+}_1$) distortion as observed from equilibrium state refinements\cite{Balachandran:2013cg,Wagner:2018kz}  (see Fig.\ \ref{Fig4}D). Since $X^{+}_5$  is also coupled to the $R^{+}_4$ and $M^{+}_3$ octahedral rotation distortions involved in M-O symmetry change at IMT, we expect that a collective structural response is needed under a dynamical structural transformation. Owing to the coupling of these modes that play a fundamental role in the structural transition, the timescale will be dictated by the slowest mode (i.e., phonon bottleneck) that is consistent with the timescale we have observed for the electronic changes. Similar behavior was seen recently for the case of layered nickelates \cite{Coslovich:2017kg} and was already known for VO$_2$\cite{Cavalleri:2004eh}. }

\section{Coherent Phonons and Nd Magnetism}

\jwf{In connection to the lattice modes that slow the CO transition, we discovered a route that connects these dynamics in the measurements of the magnetism at the Nd site in relation to Ni. Based on mean field theory, the long-range Nd sub-lattice order is induced by Ni sub-lattice  and arises from a very weak magnetic coupling to Ni via the Nd-O-Ni bonding that acts as a field to order the nominally paramagnetic Nd moments\cite{GarciaMunoz:1994dk}. This can be probed by tuning to the Nd M$_4$ resonance ($\sim$ 1000 eV) at the same wave-vector as the Ni ordering\cite{Scagnoli:2008iu}, allowing us to explore how the Nd ordering responds to the change in Ni ordering. We find that under full collapse of order at the Ni sub-lattice, the Nd sub-lattice is less disordered. A quench of Ni magnetic moment will fully quench the effective field that maintained the Nd order and a full collapse would be expected on the Nd site as well, but not observed in measurements. Furthermore, the dynamics of the Nd order show that in addition to a fast initial drop at the same rate as the collapse of Ni magnetic order (see Fig.\ \ref{Fig3}B), there is a strong oscillatory component  with a period of $\sim$ 450 fs that becomes rapidly damped within a few ps. This magnetic dynamics are generated by a coherent phonon that is strongly coupled to the Nd magnetic order. Note that if we restrict the fit of the Nd data to the early time region (between Nd intensity of 0.6 and 1) then it results in the same width of initial drop in order as the Ni case, which implies the Nd senses that the Ni long-range order collapses in the first 200 fs. Further, the extended temporal fit, including the relaxation, separates the coherent oscillation period that is consistent with the soft phonon associated with the Nd site\cite{Zaghrioui:2001kk}, which we also observed with optical reflectivity in the paramagnetic phase. These findings allows us to tie the frequency of this oscillation to a $X^{+}_5$ phonon mode that is present in both low and high temperature phases. (Figure S1). Note that with the time resolution of this experiment we are not able to see the faster modes associated with the breathing mode  ($R^{+}_1$) or rotations, which occur in the sub 100 fs regime. The ringing in the Nd magnetic scattering is related to changes in the Nd-O-Ni exchange path that is explored in more detail in the following. As noted by the dashed line in Fig.\ \ref{Fig3}, the period of the damped oscillation is consistent with the IMT transition time, implying a direct structural link with the electronic transition. Observation of phonon mode at the wavevector of the magnetic order demonstrates a coherent coupling mechanism between a structural distortion and magnetism, which was anticipated in PrNiO$_3$ and NdNiO$_3$\cite{Park:2012hg,Varignon:2017is,Mercy:2017il}, but not previously observed.}

\begin{figure}[t]
\centering
\includegraphics[width=.45\textwidth]{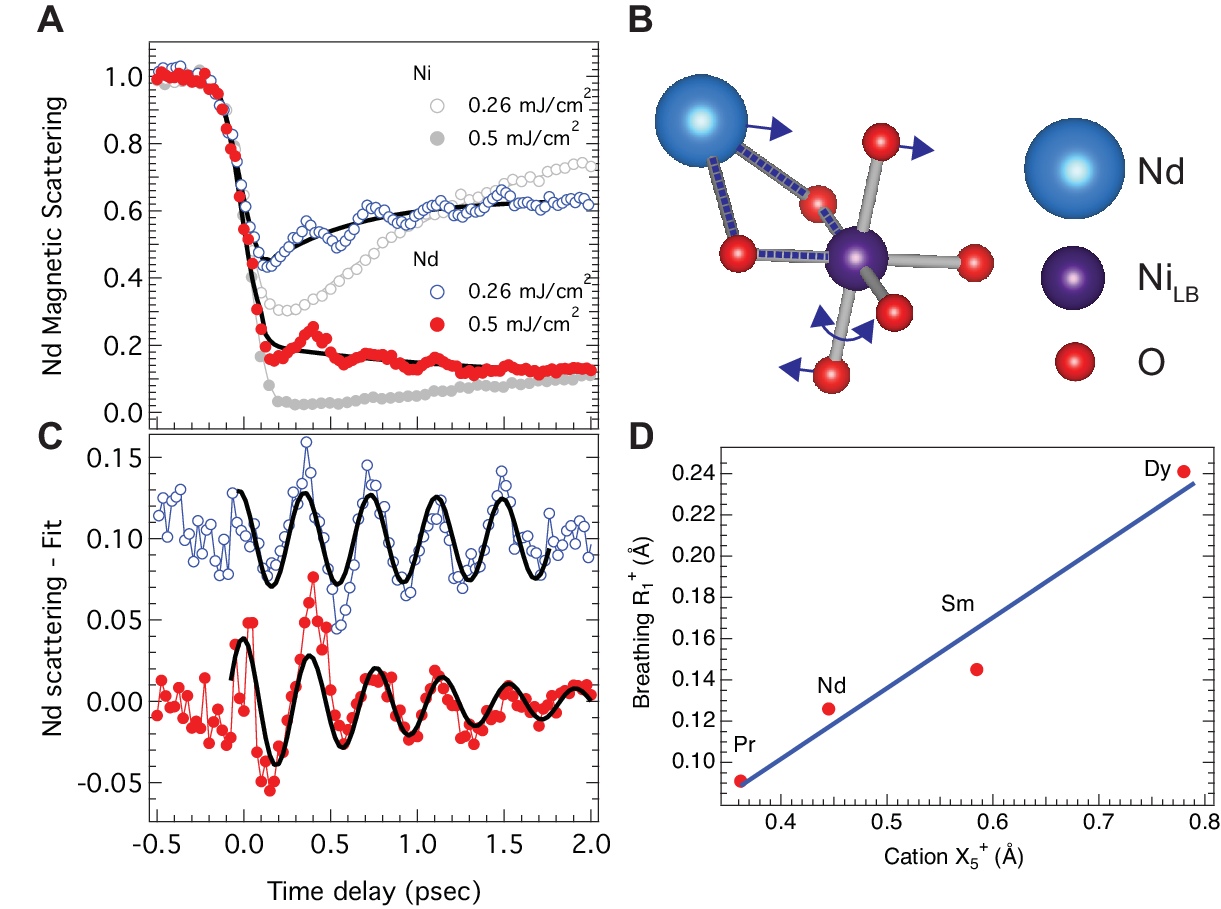} 
\caption{\label{Fig4} {\bf Coherent lattice contributions to Nd order dynamics}  (A) Nd magnetic scattering as a function of fluence in direct comparison to the Ni magnetic response. The statistical error is the order of the point size. (B) An illustration of the Nd-O soft mode associated with a coordinated Nd-O bond changes and NiO$_6$ rotation.  (C) Difference between the fit and data to extract the time period of the coherent phonon (black lines). (D) Correlation between the magnitudes of the static breathing($R^{+}_1$) and A-cation modes($X^{+}_5$).}
\end{figure}

\jwf{To better understand this coherent phonon mode and how it changes the induced Nd magnetic order\cite{Matsuda:1990fg,Scagnoli:2008iu}, we utilized theoretical calculations (see supplement) and insight from calculations of Nd-TM coupling for the case of NdFeO$_3$\cite{Chen:2012bf} as no such calculations exist for \NNO. Figure\ \ref{Fig4}A, shows a portion of the unit cell with the Nd atom and the shortest exchange pathway to the high-moment Ni$_{LB}$ site. Calculations show that this $\sim$80 cm$^{-1}$ A$_g$ symmetry mode involves the motion of the Nd atom that also drives a tilting/rotation of the NiO$_6$ octahedra without any motion of the Ni atom. Since the exchange, $J_{Nd-Ni}$, is directly connected with the Nd-O bond length, this can provide a direct connection between the magnetic exchange and the lattice vibrations. If the bond length changes, $J_{Nd-Ni}$ varies and the induced Nd order varies correspondingly. Nevertheless, early steady state estimations of the $J_{Nd-Ni}$ from powder samples have placed its value at $\sim$0.04 meV\cite{Matsuda:1990fg}  suggesting a slow response based on this energy scale. This  indicates that a simple picture may not applicable and more extensive non-equilibrium many-body interactions are dominating the ultrafast IMT. To look at this coupled mode in more detail, we show two fluences for the measurement of the Nd dynamic magnetic response.  Both show a clear coherent mode and the fact that it has a longer lifetime at low fluence can be associated with an excitation in the insulating phase (see Fig.\ \ref{Fig4}C). To quantify this mode more directly, we will break the Nd data into two parts: a fast initial decay followed by a slow recovery and the coherent oscillation.}

\jwf{In the first 200 fs, the Nd magnetic collapse has the same timescale as the Ni magnetic data implying a connection between the two processes. Usually, the induced Nd magnetic order is considered as a paramagnetic moment in the large local field due to proximity to the ordered Ni atoms\cite{Matsuda:1990fg}, as was evidenced in resonant soft X-ray scattering at the Nd M-edge. To first order, one would accordingly expect a slow paramagnetic relaxation and not an ultrafast response. However, recent work has shown that in ErFeO$_3$ the Fe magnons hybridize partially with Er spin fluctuations\cite{Li:2018ha}, providing at least a partial pathway for the Ni magnetic excitations to couple to the Nd. This seems consistent with the observation that the fast drop in Nd order is roughly proportional to what is seen for the case of Ni. However, the connection to Ni order collapse does not explain the coherent oscillation, which was not seen for Ni and has a  longer period than expected for Ni magnons ($\sim$85 fs period). To quantify the oscillation, we take the difference between the data and a  simple fit shown in Fig.\ \ref{Fig4}B and fit a cosine to this difference, we determined the oscillation periods for both fluences to be 378 $\pm$ 10 fs, , which is close to the IMT transition time of 450 fs.. As already discussed earlier, this period is consistent with the observed phonon modes by Raman spectroscopy\cite{Zaghrioui:2001kk} and our calculations (see supplement) associated with motion of the Nd atom, which causes a coordinated increase/decrease of the shortest Nd-O-Ni bonds highlighted in figure Fig.\ \ref{Fig4}B. Since the magnetic order can be affected by the change in bond-length, this results in a modulation of the Nd magnetic order at the phonon period. }

\jwf{To understand the implications of the observed dynamics, we recall that it is well known that all of the equilibrium atomic positions in the unit cell are strongly affected by the IMT, in concert with the  A-site cation shifting to a new equilibrium position in the metallic phase\cite{GarciaMunoz:2009hm}. An abrupt change in the equilibrium atomic positions arising at the first order transition provides in turn a mechanism for the displacive excitation of phonons. Interestingly, incoherent and coherent dynamics coexist in our experimental observation. The stronger, but much faster order melting, at both Ni (complete melting) and Nd (incomplete melting) is connected to both the ultrafast lattice dynamics and details of the magnetic exchange between Nd and Ni. Deeper insight will require developing a model linking the bond-length changes to the magnetic coupling between Nd and Ni. On the other hand, across the entire phase diagram of RNiO$_3$ nickelates\cite{Balachandran:2013cg}, the magnitude of A-cation displacement ($X^{+}_5$) correlates with the magnitude of the breathing mode ($R^{+}_1$) shown in Fig.\ \ref{Fig4}D . Decreasing the size of R increases the magnitude of the distortions involved in the structural phase transition and correspondingly the temperature of the IMT. In addition, the AFM order in NdNiO$_3$ was found to contribute to the stabilization of the charge ordered state\cite{Haule:2017ft,Mercy:2017il}, which should create dynamic interplay between the two under the AFM order collapse as supported by our experiments. }

\jwf{Here, using time domain measurements, we see that the displacive excitation of Nd phonon ( $X^{+}_5$) is in phase with the ultrafast melting of the Ni order. Therefore,  the displacive mechanism suggests that as soon as the magnetism is quenched, the minimum of the Nd lattice potential shifts to a new value, which is seen in the static measurements across the IMT\cite{GarciaMunoz:2009hm}. A possible link to the lattice excitation with the Nd magnetic order collapse is pointed out by recent ultrafast demagnetization experiments in rare earth (RE) metals\cite{Frietsch:2020cp}, which demonstrated that the magnetic dynamics of 4f shell of Nd is strongly tied to the orbital momentum value of RE ion. Weak versus strong coupling to the lattice was observed in Gd (L=0) and Tb (L=3), respectively, which depends on the orbital momentum configuration of the 4f shell, indicating that the magnitude of spin-orbit coupling with a large contribution from orbital momentum mediates a proportionally stronger ultrafast coupling with the lattice. The larger orbital momentum enhances the coupling with the lattice. Although \NNO\ is in the insulating ground state at 70 K in our measurements, the same considerations are expected to apply since the larger orbital momentum of Nd$^{3+}$ ion is even larger (L=6) for 4f$^3$ configuration. This picture points out the strong coupling of the 4f shell magnetism with the lattice and connects with our observations of enhanced coherent phonon modulation of the Nd magnetic order. Moreover, under a picture of mean field magnetic interactions between Ni and Nd magnetic sub-lattices, the observed coherent oscillations imply that total angular momentum of the 4f shell oscillates at the coherent phonon frequency. In fact, the deviation of the harmonic fit from the data (Fig.\ \ref{Fig4}C) indicates the possible anharmonic nature of the observed dynamics, which is consistent with a periodic change in the angular momentum as the origin for (anharmonic) coherent modulation of the magnetization. Since our experiment cannot resolve yet the orbital and spin momentum components of the total angular momentum, these interesting points deserve further studies. The coherent phonon modulation of the total angular momentum that we observed activates the role of the spin-orbit coupling in the dynamics and the enhanced interaction with the lattice along with it\cite{Frietsch:2020cp}. At the early times of less than 200 fs, where magnetic order collapses promptly, the temporal resolution is insufficient to resolve the details of the interaction between the lattice and magnetic orders. However, since the amplitude of coherent modulation of Nd magnetic order correlates with the amplitude of the magnetic collapse fast than 200 fs and at the same time it tracks the Ni order collapse on this timescale, we infer that the cooperative dynamics between the magnetism and the lattice is extremely fast.Our measurements thus provide evidence for angular momentum exchange between magnetic order and the lattice at a frequency of a few THz, serving as direct witness for the cooperation between the magnetic and structural degrees of freedom that further establishes collectively the response of the IMT.}

\jwf{This together with the different timescales for magnetic order collapse and the IMT are consistent with a picture for \NNO where the antiferromagnetic order is the primary order that supports the CO phase. The displacive excitation, results in Nd atoms moving from the larger $X^{+}_5$ distortion in monoclinic phase to a smaller one that is characteristic for the high symmetry orthorhombic phase\cite{GarciaMunoz:2009hm}, which can induce an instantaneous enhancement of the $X^{+}_5$ as the system stabilizes to the new equilibrium value. The result of this non-equilibrium enhancement of the $X^{+}_5$ displacement is to favor the insulating state, as noted by the behavior discussed above. The first coherent oscillation of large magnitude at the Nd site is observed at a time delay which corresponds well with  respective time delay of the IMT tracked by the THz conductivity (see Fig.\ \ref{Fig3}). Such a conclusion has been suggested theoretically\cite{Park:2012hg,Varignon:2017is,Mercy:2017il}, but has been too difficult to prove experimentally given the concomitant nature of the electronic, magnetic and structural phase transitions. Here, using time-domain techniques, we were able to extract insight from the fundamental timescales and show clearly that magnetism is the driver of the IMT by revealing that the initial magnetic collapse triggers the loss of CO and subsequent IMT.  }

 \subsection{Theory of Distortion Dependent Properties}
\begin{figure}[t]
\centering 
\vspace{-3em}
\includegraphics[width=.5\textwidth]{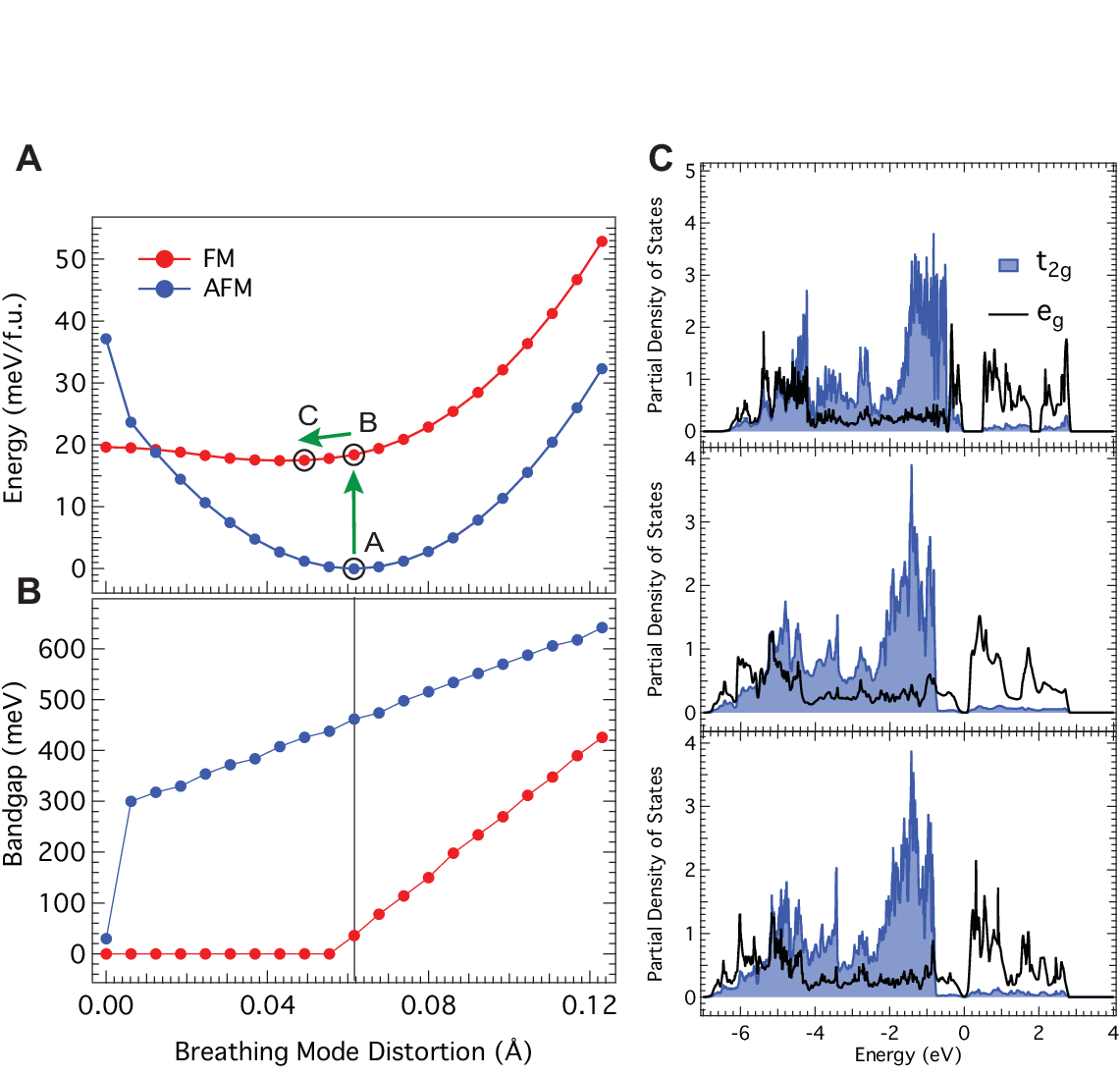} 
\caption{\label{Fig5}{ {\bf Calculated electronic and magnetic structure} (A) Dependence of AFM-E$^\prime$ and FM states as a function of breathing mode distortion\cite{Wagner:2018kz}. The two step path discussed in the text is illustrated by points A,B, and C. (B) The bandgap at as a function of the breathing mode distortion. The vertical line is at the distortion of points A and B. (C) Orbital dependent Ni density-of-states corresponding to points A, B, and C. indicated in panel A with the Fermi level set to 0\,eV. }}
\end{figure}

While a full dynamical theory for the complex \NNO\ unit cell is currently out of reach, to obtain deeper insight into how magnetic order couples to the charge order, we employ density functional theory (DFT) here to examine the electronic structure and energetics of various magnetic states as a function of static lattice distortions. The energetics of the AFM-E$^\prime$ monoclinic structure is compared against the ferromagnetic (FM) solution, which in this context serves as a proxy of the paramagnetic (PM) state, which is difficult to approximate at  the DFT level. However, note that the FM solution that is provided solely by DFT has orthorhombic symmetry and can capture well the essential structural characteristics of the structural M-O transition observed during IMT. To include the influence of correlations, a plus Hubbard $U$ = 2 eV correction was used on the Ni d orbitals, which accurately captures the details of the AFM insulating phase \cite{Varignon:2017is}.  Specifically, we investigate the evolution of the magnetic and electronic properties as a function of the breathing mode distortion (Fig.~\ref{Fig5}A). We focus on this particular distortion as it is the primary distortion active at the $Pnma \rightarrow P2_1/n$ transition; in addition,  the DFT studies indicate that Jahn-Teller and rumpling distortions are not operative in determining the magnetic order (see Figure S5). 

As a function of the cooperative breathing mode distortion, Fig.~\ref{Fig5}A plots two energy curves corresponding to insulating AFM-E$^\prime$ and ferromagnetic (FM) order\cite{Wagner:2018kz}.  The minimum for the AFM case agrees with the experimental value of the breathing mode distortion and the resulting magnetic order has $S\sim$1 on the LB site and $S=0$ on the SB site. The AFM phase is insulating at all magnitudes of breathing distortion, but the FM phase is only insulating at values of breathing distortion greater than point C (see bandgap changes versus breathing mode amplitude in Fig.\ \ref{Fig5}(B)). Although a non-collinear spin structure also satisfies the symmetry for E$^\prime$-type order \cite{GarciaMunoz:1992dj,RodriguezCarvajal:1998dy, Scagnoli:2006ja,Scagnoli:2008iu}, our results together with other recent theoretical results strongly suggest that the collinear phase is the stable ground magnetic state \cite{Varignon:2017is,Haule:2017ft}. In the calculations shown here, the energy of the FM state is always higher than the AFM state except close to zero breathing mode distortion, corresponding to the high temperature orthorhombic symmetry of the paramagnetic metallic phase. Further, we note that the gap between AFM and FM at the AFM minimum is $\sim$kT$_N$. To understand the evolution of these different degrees of freedom upon optical excitation, we consider a sequential process. 

Representative for early time delays in the experiment, the arrow from point A to B in Fig.\ \ref{Fig5}A corresponds to a direct optical excitation at a timescale faster than the lattice can respond and alter the cooperative breathing mode distortion (i.e., a shift along the horizontal axis). As shown by the density of states (DOS) in Fig.\ \ref{Fig5}B, at this point on the FM curve, the system is still insulating. Since point B is not at the minimum of the FM energy curve, the system will then move towards smaller breathing distortion and the minimum located at point C, where the system becomes metallic (see Fig.\ \ref{Fig5}C). However, note that the energy barrier between  point C and zero breathing distortion is observed to be much smaller than  kT$_N$, which indicates that our DFT calculation predicts effectively a disordered or thermally unstable breathing distortion in the orthorhombic phase. \jwf{From our analysis of the calculated density of states above, two key components change between AFM (A) and FM (B) configurations. First, the local moment on the SB site, which is $0\,\mu_B$  for the AFM phase converts to $\sim0.5\,\mu_B$ in the FM state. This arises from a local rearrangement of the spin-dependent $e_g$ orbital occupancy on the SB site. Secondly, the difference in Born effective charges between the LB and SB sites increases (see details in supplement Table S3). This is associated with the redistribution of oxygen holes in the lattice due to the change in magnetic state, similar to the process that would be triggered via light induced ICT. }

\section{Discussion}

\jwf{Above we have shown evidence for a non-thermal IMT in NdNiO$_3$ driven by the collapse of magnetic order. The loss of long range AFM involves magnetic ordering on both cations sites and occurs faster than 200 fs connected to the fundamental timescale of the magnons. On the other hand, the IMT occurs at a timescale of $\sim$450 fs, which is set by the phonon bottleneck of a Nd-O phonon mode. This phonon modulates the Nd magnetic order at 2.6 THz in the presence of the large angular momentum of the 4f shell, which in turn activates the role of spin-orbit coupling in mediating the magnetic interaction with the lattice. Strikingly, we collected direct evidence for development of magneto-structural coupling being substantially faster than IMT, demonstrating that magnetism indeed stabilizes the lattice order and can be used for ultrafast control of IMT.
Now, we can expand and discuss this in connection to a much wider class of materials with entangled order parameters where the ultrafast dynamic approach can be used to provide insights not easily accessible under equilibrium conditions. It is useful to consider the present results in the context of other ultrafast experiments not only on NdNiO$_3$ but more generally in the context of photoinduced IMT dynamics in other materials. }

\jwf{In manganites, vanadates and 3d transition metal oxides\cite{Radaelli:1995cz,Bao:1997ft,Matsubara:2007fv,Leonov:2016gm,Kalcheim:2019bc,Frandsen:2019gw,Trastoy:2020di}, where the IMT has a Mott character and occurs proximal to a change in magnetic and structural order, similar opportunities arise for exploiting systems dominated by strong coulomb interactions. The photoinduced inter-site charge and spin transfer can be used to exploit ultrafast magnetic responses to investigate their dynamical coupling with either IMT or with the systems structural degrees of freedom. As our studies show, the rare earth cations can be used as a tool to enhance the observations of spin-lattice interaction mediated by larger spin-orbit coupling present within their 4f shells. Furthermore, the magnetic dynamics influencing the structure can be enhanced further at the B-site cations by spin-orbit (SO) coupling arising in 5d transition metal oxides like in osmates and iridates\cite{Mandrus:2001ia,Calder:2012jf,Zheng:2016gr,Lu:2017fja}. In these systems, the SO couplingcompetes with Coulomb electronic interactions and Slater IMT are observed\cite{Mandrus:2001ia,Calder:2012jf}. We also found that the dynamical approach for disentangling degrees of freedom at the IMT can become potentially useful for other compounds like FeS and MnB$_4$ for which spin-phonon instabilities were identified already\cite{Liang:2014en,Bansal:2020dl}. We accordingly expect that the multimodal ultrafast approach for dynamically establishing the role of different degrees of freedom at IMT will become very useful in the years to come. Fundamentally, we can use such studies to understand the impact of magnetism in controlling the conductive properties of quantum materials with large figures of merit. }

\section{Conclusion}

\jwf{In summary, using X-rays to disentangle the electronic and magnetic degrees of freedom has provided a concise picture of how light interacts with strongly correlated matter. By tracking the distinct timescales one can gain not only a clearer picture of how charge, magnetic, orbital, and lattice orders evolve, but also which are key to driving the transition and which are triggered as a response to the changing fundamental order parameter. Using this approach, not only can we see that magnetic order is the fundamental order parameter for NdNiO$_{3}$ that stabilizes the insulating phase in cooperation with charge order, but we also provide a paradigm to unravel entangled order parameters in many complex materials of contemporary interest. We thus demonstrated that magnetism  can be manipulated in a TMO to provide access to ultrafast control of metal-insulator transitions. These studies also revealed direct evidence for coherent phonon coupling to magnetic order during the photoinduced IMT, substantiating that a coherent interaction between lattice and magnetic degrees of freedom can be controlled on ps timescales and at THz speeds. }

\section{Acknowledgements}
 \begin{acknowledgements}   J.W.F. wants to acknowledge the help with science and data analysis from N. Laanait. V.S., J.Z., R.D.A, J.C., H.W., J.M.R, and J.W.F were supported by the Department of Energy grant DE-SC0012375 for work with ultrafast X-Ray and optical experiments and analysis of the data with theoretical support. D.P.\ was supported by the Army Research Office (Grant No.\ W911NF-15-1-0017). J.C. was also supported by DOD-ARO under Grant No. 0402-17291 and by  the Gordon and Betty Moore Foundation's EPiQS Initiative through Grant No. GBMF4534. Work at the Advanced Photon Source, Argonne was supported by the U.S. Department of Energy, Office of Science under Grant No. DEAC02-06CH11357. Use of the Linac Coherent Light Source (LCLS), SLAC National Accelerator Laboratory, which is a DOE Office of Science User Facility, under Contract No. DE-AC02-76SF00515.  Electronic structure calculations were performed using computational resources provided by the DOD-HPCMP. J.W.F. would like to acknowledge many insightful conversations with A.J. Millis and D. Khomskii.
 \end{acknowledgements}

\bibliography{NdNiO3_Ultrafast_PRB_final_v1.bib}

\end{document}